\documentclass[prc,twocolumn,showpacs,preprintnumbers,amsmath,amssymb]{revtex4}

\usepackage{amsmath,amssymb,multirow,epsfig,bm}

\newcommand{\beq}{\begin{equation}}
\newcommand{\eeq}{\end{equation}}
\newcommand{\bea}{\begin{eqnarray}}
\newcommand{\eea}{\end{eqnarray}}

\newcommand{\benn}{\begin{displaymath}}
\newcommand{\eenn}{\end{displaymath}}

\newcommand{\angstr}{\rm\AA}

\begin{document}

\title{Magnetism in doped two-dimensional honeycomb structures of III-V binary compounds.}

\author{Krzysztof Zberecki$^1$}
\affiliation{$^1$Faculty of Physics, Warsaw University of Technology, ul. Koszykowa 75, 00-662 Warsaw, Poland}
\email{zberecki@if.pw.edu.pl}
\date{\today}

\begin{abstract}
 Using first-principles plane-wave calculations systematic study of magnetic properties of doped two-dimensional honeycomb structures
 of III-V binary compounds have been conducted, either for magnetic or nonmagnetic dopants. Calculations show, that all cases 
 where magnetic moment is non-zero are energetically more favorable. For such cases band structure and (partial) density 
 of states were calculated and analyzed in detail. The possible applications of these structures were also discussed.\\
\end{abstract}


\maketitle

\section{INTRODUCTION}
 Since its discovery in 2004 graphene ~\cite{nov1} draws much attention beacuse of uniqe features of this two-dimensional system. 
 Graphene is composed of a sp$^ {2}$-bonded carbon atoms forming honeycomb structure. It has very interesting electronic structure 
 with characteristic, linear energy dispersion near K point  of Brilloin zone. Summary of the subject can be found for example in 
~\cite{graph1}. \newline
  Shortly after, experimental techniques allowed fabrication of other new two-dimensional materials, like BN and MoS$_2$ honeycomb 
 structures ~\cite{nov2} or ZnO monolayers ~\cite{zno1}. The discovery of such stable two-dimensional material like graphene 
 triggered search for similar structures made from different compounds. Up to now many of these hypotetical structures 
 constructed from silanene (2D Si) and germanene (2D Ge) ~\cite{sil1, sil2}, III-V compounds ~\cite{III-V1}, SiC ~\cite{sic1} 
 or ZnO ~\cite{zno2} have been studied theoretically. \newline
 On the other hand graphene and other nano-scale materials are recognized as future building blocks of new electronics technologies
 ~\cite{nano1}, including spintronics (e.g. ~\cite{spin1}). In the case of low (one- and two-) dimensional structures  problem arises
 because of famous Mermin-Wagner theorem ~\cite{mermin1}, which prevents ferro- or antifoerromagnetic order to occur in finite 
 temperatures, which is essential for spintronics and other modern applications. This started the theoretical  and experimental 
 search for magnetism in graphene and other structures. One of the most promising directions is emergence of magnetism 
 in such structures as an effect of presence of local defects ~\cite{exp1}. According to works of Palacios et al. ~\cite{palacios1} 
 and, independently, of Yazev ~\cite{yazyev1} single-atom defects can induce ferromagnetism in graphene based materials. 
 In both cases, the magnetic order arises as an effect of presence of single-atom defects in combination with a sublattice
 discriminating mechanism, in agreement with Lieb's theorem ~\cite{lieb1}. Based on these findings several theoretical studies have 
 been conducted in search for magnetism in low-dimensional structures either for graphene and BN ~\cite{bn1} or other (hypothetical)
 structures like SiC ~\cite{sic1}. \newline
 In this paper influence of local defects on magnetic structure of two-dimensional honeycomb structures of GaN, AlN and InN have been
 analysed by means of $ab$-$initio$ calculations. Since bulk versions of these compounds are very important semiconductors in todays 
 electronics it would be interesting to check whether such two-dimensional materials could have non-zero magnetic moment. Despite of 
 the fact that neither of them have been yet synthesized, calculated cohesion energies ~\cite{III-V1} suggest that such structures
 would be stable and their experimental procurement is highly probable. \newline
 Next section contains computational details followed by results. Last section concludes this work.

\section{COMPUTATIONAL DETAILS}
To investigate magnetic properties of GaN, AlN and InN honeycomb structures a series of $ab$-$initio$ calculations have been conducted 
with use of DFT VASP code ~\cite{vasp1,vasp2} with PAW potentials ~\cite{vasp3}. For both spin-unpolarized and spin-polarized cases 
exchange-correlation potential has been approximated by generalized gradient approximation (GGA) using PW91 functional ~\cite{pw91}. 
Kinetic energy cutoff of 500 eV for plane-wave basis set has been used. Supercells of size 4x4x1 have been checked to be large enough 
to prevent defects interact with its periodic image. In all cases for self-consistent structure optimizations, the Brillouin zone (BZ) 
was sampled by 20x20x1 special k points. All structures have been optimized for both, spin-unpolarized and spin-polarized cases unless 
Feynman-Hellman forces acting on each atom become smaller than 10$^{-4}$ eV/$\angstr$. A vacuum spacing of 12 \angstr \ was applied 
to hinder the interactions between monolayers in adjacent supercells. Calculated lattice constants are in agreement with ~\cite{III-V1}.

\section{RESULTS}
\begin{figure}[h]
\vspace*{-0.3cm}
\includegraphics[scale=0.3,angle=-90]{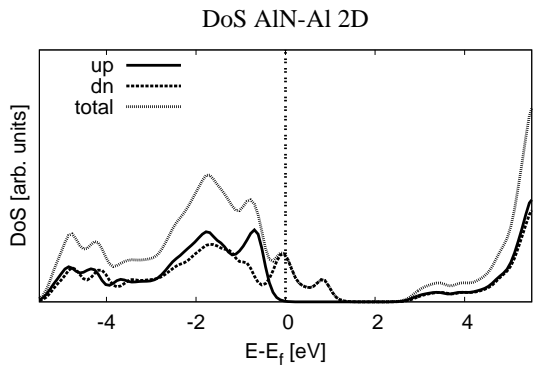}
\vspace*{0cm}
\caption{Density of states of Al-vacant AlN.} 
\label{fig1}
\vspace*{-0.4cm}
\end{figure}

As mentioned, non-magnetic honeycomb sheets can attain spin polarized states due to presence of local defects. In this work two kinds 
of defects have been analysed - vacancies and subsitutions. \newline
For all three compounds a vacancy was generated first by removing a single atom, Al, Ga, In or N from each supercell, then the atomic
structure was optimized. In all cases structures with single N vacancy are non-magnetic, while Al, Ga or In vacancies induce non-zero 
magnetic moment, equal to 3.00 $\mu_{B}$. This is in disagreement with Lieb's theorem, which states that magnetic moment should be 
equal to 1.00 $\mu_{B}$ when one of sublattices has exacly one atom more/less than the other (e.g. N$_{Al}$ - N$_{N}$ = $\pm$ 1). This discrepancy
can be addressed to charge transfer from Al(Ga,In) to N. 
Fig. \ref{fig1} shows density of states (DoS) for spin-polarized Al-vacant AlN,
on which difference between majority spin (up) and minority spin (dn) in the vicinity of Fermi level (horizontal line)
can be observed which is the main source of non-zero magnetic moment. Analysis of calculated partial magnetisation shows that almost all 
magnetic moment is situated on p-states of N atoms located in the area of vacancy. This is in full agreement with previous studies of 
vacancies in SiC ~\cite{sic1}. \newline

In the case of substitution the procedure was as follows. For all three compounds various single foreign atoms have been substituted, then 
structure has been optimized. In the case of AlN and GaN, Al or Ga has been subsituted by atoms from 4-th period of periodic table 
from K to Zn, including Na and Mg for AlN. In the case of InN, In has been subsituted by atoms from 5-th period of periodic table from Rb to Cd, excluding Tc. 
In all three compounds, N has been subsituted by C, B and P atoms. Table I shows calculated magnetic moments and differences between total energy of spin-unpolarized
and spin-polarized states, $\Delta$E = E$_{nsp}$ - E$_{sp}$, i.e. positive value of $\Delta$E means that spin-polarized state is more 
energetically favored. This is the case for all compounds with non-zero magnetic moment - the largest energy differences (up to 1.25 eV) are for
instances with highest values of magnetic moment. 
For AlN highest values of induced magnetic moment are in case of Mn, Co (4.00 $\mu_{B}$) and for Fe (4.26 $\mu_{B}$). 
Similar situation can be observed for GaN doped with Mn, Co and Fe. With decreasing number of d-shell electrons
value of magnetic moment drops as well as for the case of Zn which has d-shell closed. In the case of InN this tendency holds although
values of magnetic moments are much smaller, being the highest for Ru and Rh. Figs. \ref{fig2a} and \ref{fig2b} show mechanism of generation of magnetic moment 
in the case of GaN doped with Ni. Left plot of Fig. \ref{fig2a} shows bandstructure of Ni-doped GaN vs. undoped one (which is a semiconductor with bandgap
equal to 2.30 eV, calculated within GGA) in spin-unpolarized case. 
One can see the formation of doping bands in the vicinity of Fermi level. The top right plot shows bandstructure of GaN+Ni in the 
spin-polarized case, where can be observed quite large splitting of these bands between spin up and down bands. 

\begin{figure}[h]
\vspace*{-0.3cm}
\includegraphics[scale=0.3,angle=-90]{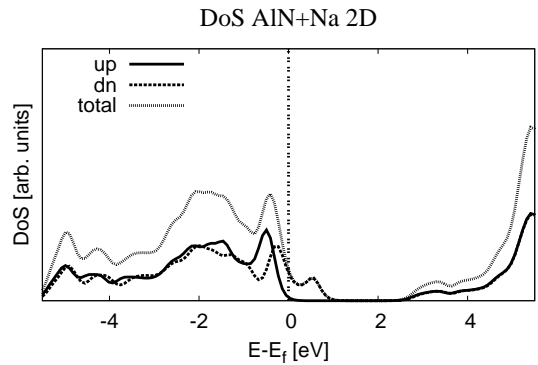}
\vspace*{0cm}
\caption{Density of states of Na-doped AlN.} 
\label{fig3}
\vspace*{-0.4cm}
\end{figure}

\begin{figure}[H]
  \begin{center}
    \begin{tabular}{cc}
      \resizebox{80mm}{!}{\includegraphics[angle=270]{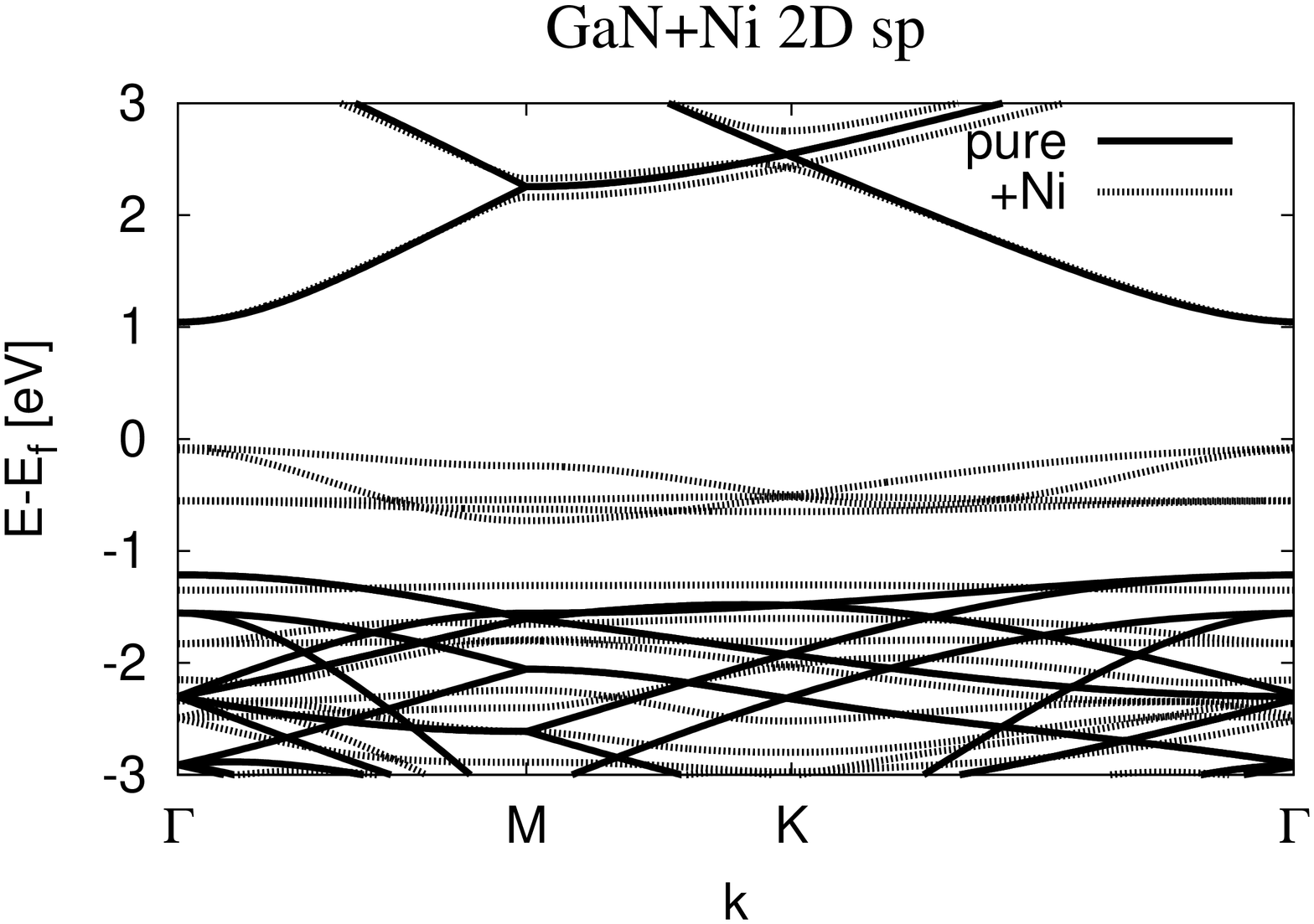}} &
      \resizebox{80mm}{!}{\includegraphics[angle=270]{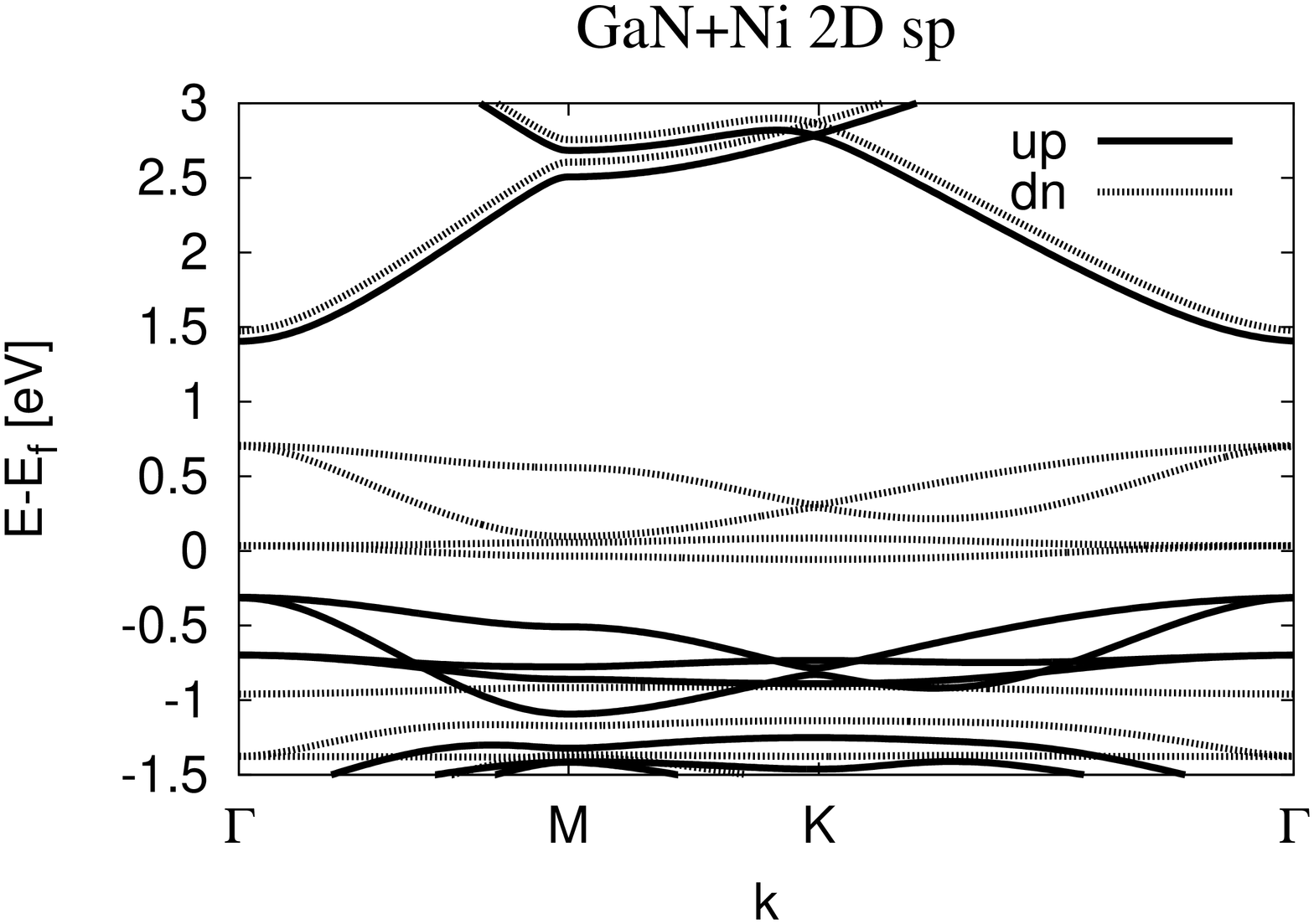}} \\
    \end{tabular}
    \caption{Bandstructure of Ni-doped GaN. Details in text.}
    \label{fig2a}
  \end{center}
\end{figure}

\begin{figure}[H]
  \begin{center}
    \begin{tabular}{cc}
      \resizebox{80mm}{!}{\includegraphics[angle=270]{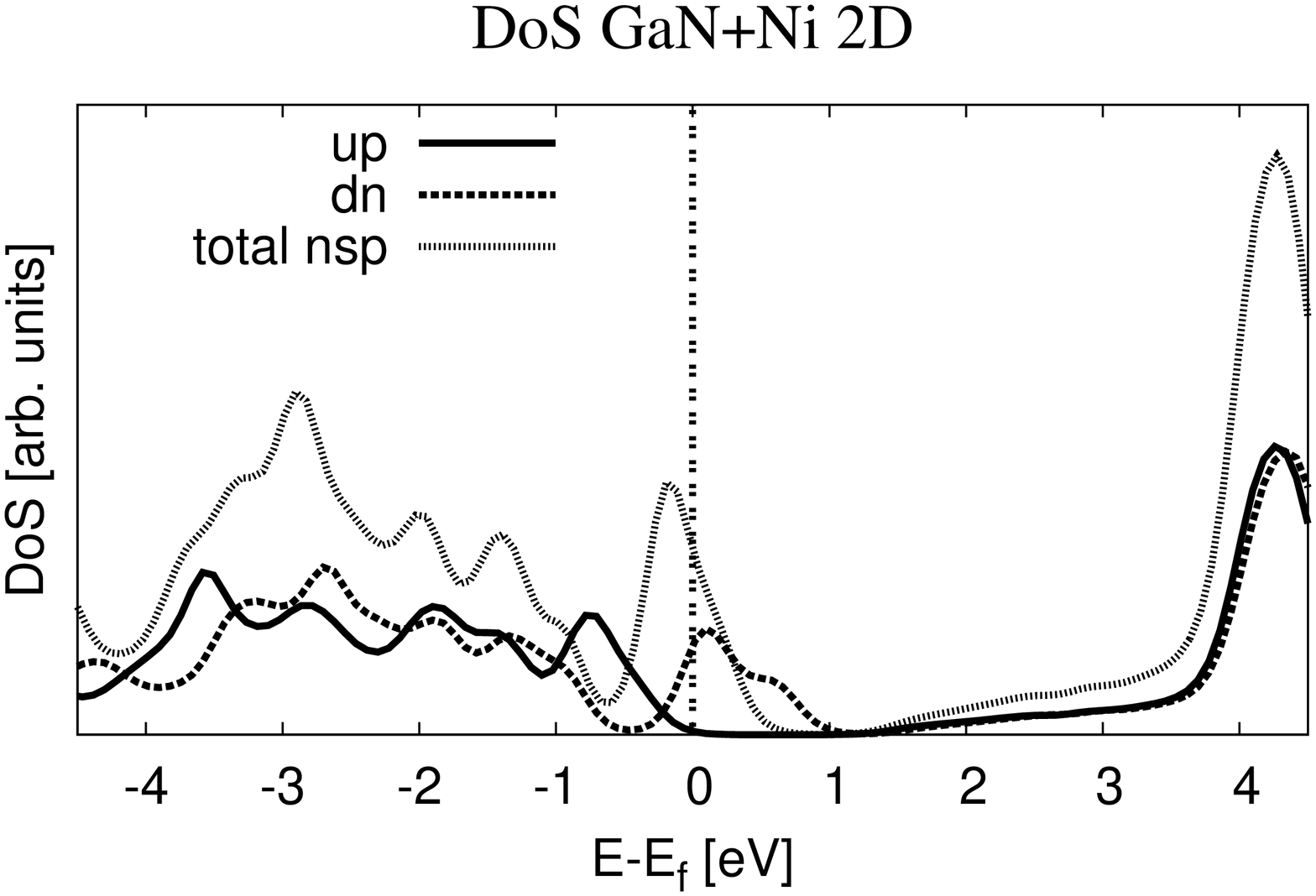}} &
      \resizebox{80mm}{!}{\includegraphics[angle=270]{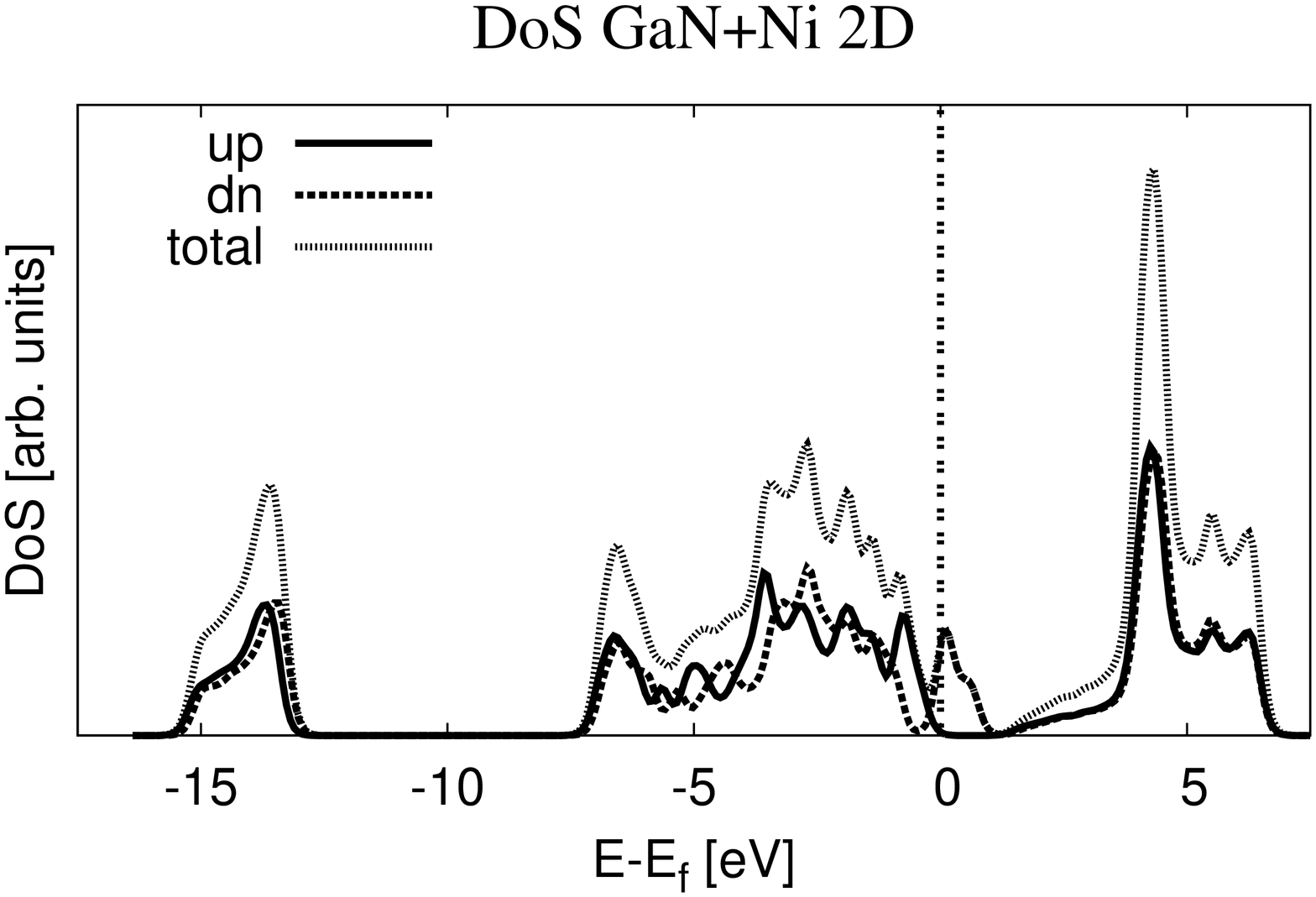}} \\
			 \resizebox{80mm}{!}{\includegraphics[angle=270]{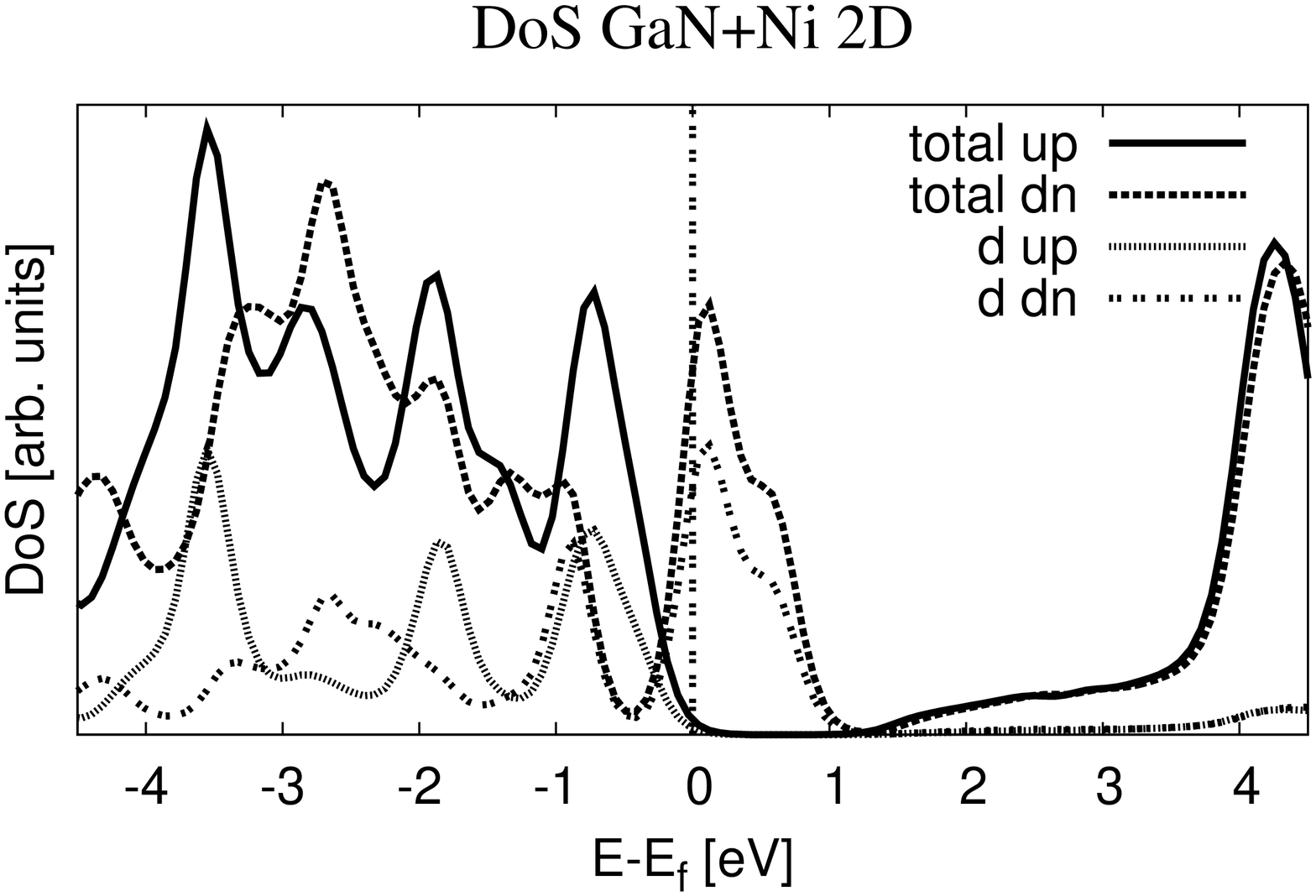}} &
    \end{tabular}
    \caption{Density of states for Ni-doped GaN. Details in text.}
    \label{fig2b}
  \end{center}
\end{figure}

Top plots of Fig. \ref{fig2b} show density of states  for spin polarized case. Right one shows total DoS vs. DoS of spin up and down. Fermi energy is almost 
exactly in the middle of splitted up and down DoS. Left plot shows vicinity of Fermi energy more closely, where 
large splitting of up and down DoS can be observed. Since almost all electrons occupying vicinity of Fermi level are d-shell electrons, which can be read from 
partial density of states (bottom left plot) the mechanism of magnetic moment emergence becomes clear. Calculations show that this mechanism is universal for all
structures having non-zero magnetic moment doped with transition metal elements. 
In case of doping with alkali metal elements and alkaline earth metal elements only in Na- and K-doped AlN calculations show non-zero magnetic moment 
(1.88 and 1.70 $\mu_{B}$, respectively for Na and K). Mechanism of formation of magnetic moment is similar to the case of vacant structures (since
Na and K have only one valence electron). Fig. \ref{fig3} shows DoS for spin-polarized Na-doped AlN, on which difference between spin up and spin down 
in the vicinity of Fermi level can be observed. 

This is similar to situation depicted on Fig. \ref{fig1} altough in the case of Na-doped structure splitting is smaller.  \newline
In case of substitution of N atom by C, B and P, only C-doped structures had non-zero magnetic moment, which was equal to
1.00 $\mu_{B}$ in all compounds. Magnetic states in all cases are lower by about 0.1 eV than nonmagnetic states. 

\begin{table}[H]
\begin{tabular}{lcccccccccccccc}
\hline

\hline
AlN &  &  &  &  &  &  &  &  &  &  &  &  & &\\ \hline
for Al            & Na & Mg & K & Ca & Sc & Ti & V & Cr & Mn & Fe & Co & Ni & Cu & Zn  \\ 
$\mu (\mu_{B})$ & 1.88 & 0.00 &1.70 & 0.00 & 0.00 & 0.96 & 0.00 & 2.98 & 3.99 & 4.26 & 3.96 & 2.98 & 1.85 & 0.00 \\ 
$\Delta$E (eV)  & 0.06 & 0.00 &0.04 & 0.00 & 0.00 & 0.11 & 0.00 & 0.70 & 0.98 & 1.09 & 0.94 & 0.38 & 0.06 & 0.00 \\ \hline
GaN &  &  &  &  &  &  &  &  &  &  &  &  & &\\ \hline
for Ga          &  &  & K & Ca & Sc & Ti & V & Cr & Mn & Fe & Co & Ni & Cu & Zn   \\ 
$\mu (\mu_{B})$ &  &  & 0.00 & 0.00  & 0.00 & 0.90 & 1.72 & 2.99 & 3.99 & 4.51 & 3.97 & 2.97 & 1.32 & 0.00   \\ 
$\Delta$E (eV)  &  &  & 0.00 & 0.00  & 0.00 & 0.12 & 0.07 & 0.66 & 1.15 & 1.25 & 0.91 & 0.30 & 0.01 & 0.00   \\ \hline
InN &  &  &  &  &  & &  &  &  &  &  &  &  & \\ \hline
for In          &  &  & Rb & Sr & Y & Zr & Nb & Mo & Tc & Ru & Rh & Pd & Ag & Cd   \\ 
$\mu (\mu_{B})$ &  &  & 0.00 & 0.00 & 0.00 & 0.00 & 0.99 & 0.39 &   -  & 1.38 & 1.91 & 0.42 & 0.00 & 0.00   \\ 
$\Delta$E (eV)  &  &  & 0.00 & 0.00 & 0.00 & 0.00 & 0.16 & 0.01 & - & 0.04 & 0.10 & 0.01 & 0.00 & 0.00   \\  \hline
\end{tabular}

\caption{Magnetic moments and total energy differences between spin up and spin down states for different transition metal elements dopants.}  
\end{table}

\begin{table}[H]
\begin{tabular}{lccccc}
\hline

\hline
AlN &  &  &  &   \\ \hline
for N            & B & C & P &    \\ 
$\mu (\mu_{B})$ & 0.00 & 1.00 & 0.00 &   \\ 
$\Delta$E (eV)  & 0.00 & 0.14 & 0.00 &   \\ \hline
GaN &  &  &  &   \\ \hline
for N            & B & C & P &    \\ 
$\mu (\mu_{B})$ & 0.00 & 1.00 & 0.00 &    \\ 
$\Delta$E (eV)  & 0.00 & 0.11 & 0.00 &    \\ \hline
InN &  &  &  &   \\ \hline
for N            & B & C & P &     \\ 
$\mu (\mu_{B})$ & 0.00 & 1.00 & 0.00 &   \\ 
$\Delta$E (eV)  & 0.00 & 0.10 & 0.00 &   \\ \hline
\end{tabular}

\caption{Magnetic moments and total energy differences between spin up and spin down states for different dopants.}  
\end{table}

\section{CONCLUSIONS}
$Ab$-$initio$ calculations have been conducted for vacancy and subsitution defects in honeycomb AlN, GaN and InN compounds.
Calculations show that in all three compounds vacancy of Al, Ga or In respectively gives magnetic moment of 3.00 $\mu_{B}$,
which is interesting conclusion from application point of view. On the other hand substitution of Al or Ga by transition metal
elements (Mn, Fe, Co) can give even higher value of magnetic moment (4.00 $\mu_{B}$). Since techniqe of implantation of metal
atoms into 2D surface has been recently reported ~\cite{nano2}, it is also very promising direction. 
On the other hand substitution by non-metallic atoms or substitution of nitrogen atoms by IV or V group atoms does not give 
significant magnetic moment. Thes results may give a hint for experimentalists seeking for two-dimensional magnetic materials. 

\begin{acknowledgments}
Numerical calculations were performed at the Interdisciplinary Centre for Mathematical and Computational Modelling (ICM) at Warsaw University.
\end{acknowledgments}

\end{document}